\documentclass[fleqn,usenatbib]{mnras}

\usepackage{newtxtext,newtxmath}

\usepackage[T1]{fontenc}
\usepackage{ae,aecompl}

\usepackage{float}

\newcommand{\hkpc}{{\ifmmode{h^{-1}{\rm {kpc}}}\else{$h^{-1}{\rm{kpc}}$}\fi}}
\newcommand{\hMpc}{{\ifmmode{h^{-1}{\rm {Mpc}}}\else{$h^{-1}{\rm{kpc}}$}\fi}}
\newcommand{\hMsun}{{\ifmmode{h^{-1}{\rm {M_{\odot}}}}\else{$h^{-1}{\rm{M_{\odot}}}$}\fi}}
\newcommand{\Msun}{{\ifmmode{{\rm {M_{\odot}}}}\else{${\rm{M_{\odot}}}$}\fi}}

\newcommand{\Sec}[1]{Section~\ref{#1}}

\newcommand{\Fig}[1]{Fig.~\ref{#1}}

\usepackage{graphicx}	        
\usepackage{amsmath, bm}        
\usepackage{multirow}           
\usepackage[dvipsnames]{xcolor} 
\usepackage{subfig}

\newcommand{\code}[1]{\texttt{#1}}

\title[The gas disruption of infalling objects]{The Three Hundred Project: The gas disruption of infalling objects in cluster environments}

\author[Mostoghiu et. al]{\parbox{\textwidth}{
Robert Mostoghiu$^{1,3,4}$\thanks{robert.mostoghiu@uam.es},
Jake Arthur$^{4},$
Frazer R. Pearce$^{4},$
Meghan Gray$^{4},$
Alexander Knebe$^{1,2,3},$
Weiguang Cui$^{5},$
Charlotte Welker$^{6,7},$
Sof\'ia A. Cora$^{8,9},$
Giuseppe Murante$^{10},$
Klaus Dolag$^{11,12},$
Gustavo Yepes$^{1,2}$
}
\vspace{0.4cm}
\\
\parbox{\textwidth}{
$^{1}$Departamento de F\'isica Te\'{o}rica, M\'{o}dulo 15, Facultad de Ciencias, Universidad Aut\'{o}noma de Madrid, E-28049 Madrid, Spain\\
$^{2}$Centro de Investigaci\'{o}n Avanzada en F\'isica Fundamental (CIAFF), Facultad de Ciencias, Universidad Aut\'{o}noma de Madrid, 28049 Madrid, Spain\\
$^{3}$International Center for Radio Astronomy Research, University of Western Australia, 35 Stirling Highway, Crawley, Western Australia 6009, Australia\\
$^{4}$School of Physics \& Astronomy, University of Nottingham, Nottingham NG7 2RD, UK\\
$^{5}$Institute for Astronomy, University of Edinburgh, Royal Observatory, Edinburgh EH9 3HJ, United Kingdom\\
$^{6}$Department of Physics and Astronomy, Zanvyl Krieger School for Arts \& Sciences, The Johns Hopkins University,  Baltimore, MD 21218, USA\\
$^{7}$Department of Physics and Astronomy, McMaster University, Hamilton, Ontario, Canada\\
$^{8}$Instituto de Astrof\'isica de La Plata (CCT La Plata, CONICET,UNLP), Observatorio Astron\'omico, Paseo del Bosque, B1900FWA, La Plata, Argentina\\
$^{9}$Facultad de Ciencias Astron\'omicas y Geof\'{\i}sicas, Universidad Nacional de La Plata (UNLP), Observatorio Astron\'omico, Paseo del Bosque, B1900FWA La Plata, Argentina\\
$^{10}$I.N.A.F. Trieste Observatory, Via Tiepolo 11, 34143 Trieste, Italy\\
$^{11}$University Observatory Munich, Scheinerstra{\ss}e 1, D-81679 Munich, Germany\\
$^{12}$Max-Planck-Institut fur Astrophysik (MPA), Karl-Schwarzschild Stra{\ss}e 1, D-85748 Garching bei M\"{u}nchen, Germany
}}

\date{Accepted XXX. Received YYY; in original form ZZZ}

\pubyear{2019}

\begin{document}
\label{firstpage}
\pagerange{\pageref{firstpage}--\pageref{lastpage}}
\maketitle

\begin{abstract}
We analyse the gas content evolution of infalling haloes in cluster environments from \textsc{The Three Hundred} project, a collection of 324 numerically modelled galaxy clusters. The haloes in our sample were selected within $5R_{200}$ of the main cluster halo at $z=0$ and have total halo mass $M_{\rm{200}}\geq10^{11}\hMsun$. We track their main progenitors and study their gas evolution since their crossing into the infall region, which we define as $1-4R_{200}$. Studying the radial trends of our populations using both the full phase space information and a line-of-sight projection, we confirm the Arthur et al. (2019) result and identify a characteristic radius around $1.7R_{200}$ in 3D and at $R_{200}$ in projection at which infalling haloes lose nearly all of the gas prior their infall. Splitting the trends by subhalo status we show that subhaloes residing in group-mass and low-mass host haloes in the infall region follow similar radial gas-loss trends as their hosts, whereas subhaloes of cluster-mass host haloes are stripped of their gas much further out. Our results show that infalling objects suffer significant gaseous disruption that correlates with time-since-infall, cluster-centric distance and host mass, and that the gaseous disruption they experience is a combination of subhalo pre-processing and object gas depletion at a radius which behaves like an accretion shock.
\end{abstract}




\section{Introduction} \label{sec:introduction}
Our current understanding of structure formation, the $\Lambda$CDM paradigm, describes the growth of structures via continuous merging of lower mass systems into more massive, denser haloes \citep{White78}. As galaxies fall into these objects, they are affected by several mechanisms due to their local environment \citep[e.g.][and references therein]{Boselli06}. Studies have shown that correlations between galaxy properties and environment become stronger in more extreme environments, such as within the virial radius of galaxy clusters. For instance, compared to the field, in high density environments early-type morphologies are more abundant \citep{Dressler80}, galaxies are redder \citep{Baldry06}, and star formation is suppressed \citep{Gomez03}.

Determining the extent of galaxy disruption in “sub-dominant” environments, such as filaments and group-sized haloes, before an in-falling galaxy passes beyond the virial radius of a galaxy cluster, also known as
pre-processing \citep{Fujita04}, is likewise important. Pre-processing has been suggested as a possible cause for recent observational results that find suppressed star formation \citep{Lu12,Cybulski14,Haines15}, lower disc fractions \citep{Roberts17} and higher red fractions \citep{Just19} in infalling galaxies at high cluster-centric radii compared with the field. More specifically, observations have shown that subhaloes, containing satellite galaxies, show signs of pre-processing \citep{Hou14} at high cluster-centric radii. In fact, it has been suggested that the quenching of satellite star formation is responsible for the majority of all quiescent (red-sequence) galaxies at $M_{\rm{star}}$ < $10^{10}$ M$_{\odot}$ by $z = 0$ and pre-processing by groups dominates this quenching \citep{Wetzel13}.

Using SDSS data, \citet{Wetzel12} and \citet{Wetzel13} showed that on first infall into a host, satellites are unaffected by the host environment and are able to actively form stars for $2-4$ Gyr after their infall. Results from \citet{Wetzel13} also imply that the satellite quenching time does not depend on the mass of the host, but rather the mass of the satellite. In contrast, \citet{Roberts17} showed that the largest degree of pre-processing was found in the smallest satellite galaxies and the largest hosts. Theoretical work by \citet{Bahe15} agrees with \citet{Wetzel13} in that there is a delay in quenching satellites, once accreted onto a host. However, they also go on to corroborate the results of \citet{Roberts17}, by showing that quenching timescales do depend on host mass. Further theoretical studies have also tried to assess how prevalent pre-processing is. Using semi-analytic models, \citet{McGee09} and \citet{DeLucia12} found that up to $\sim 40$ per cent of galaxies residing in clusters at z = 0 had previously spent a significant time in group environments. However, using a similar technique, \citet{Berrier09} found that pre-processing was a secondary process in galaxy evolution. Using a semi-analytic model, \citet{Cora19} found quenching times consistent with the ones presented in \citet{Wetzel13}, and a star formation quenching of low-mass satellites (i.e.  $M_{\rm{star}}$ $\sim$ $10^{10}$ M$_{\odot}$) that supports the \citet{Wetzel13} `delay-then-rapid' scenario. However, they showed that this scenario does not accurately describe the quenching experienced by the $z=0$ passive satellites with higher stellar mass, as the duration of both phases is of the same order of magnitude. Thus, they suggest a `delay-then-fade' quenching scenario that accounts for quenching processes in which both phases have comparable time-scales.

The missing ingredient from these studies is an extensive examination of the gas in pre-processed objects. \citet{Bahe13} made some attempt to quantify the gas stripping beyond the cluster virial radius by using the instantaneous gas fractions of infalling objects around clusters. Their main finding showed that radial gas fractions decrease with decreasing cluster-centric distance, and contamination from backsplash and pre-processed galaxies (which will also increase with decreasing cluster-centric distance) brings the distribution down. This is in disagreement with \citet{Lotz19} and \citet{Arthur19}, who find that the instantaneous gas fraction does decline radially, but the gas in (sub)haloes is lost on first passage, rather than contamination being the cause for the radial decline. In fact, in \citet{Arthur19} it was postulated that the majority of gas in infalling objects is stripped by some sort of accretion shock at $1.5-2R_{200}$, where $R_{200}$ is the radius of the halo at 200 times the critical density of the universe at that redshift. However, whilst galaxy star formation quenching has been well studied, a full examination of the gas in infalling objects is needed in order to possibly alleviate tension in the literature and learn more about pre-processing.

In this work we use the \textsc{The Three Hundred} dataset \footnote{\url{https://the300-project.org}}, a sample of over 300 galaxy clusters simulated with full-physics hydrodynamics out to $> 5R_{200}$ of each galaxy cluster \citep{Cui18}, to assess the level of gaseous disruption of haloes and subhaloes beyond $R_{200}$. We extend the $z=0$ analysis done in \citet{Arthur19} by using the orbital histories of $> 10^5$ infalling (sub)haloes in a range of environments to quantify gas-loss since infall, which we arbitrarily define as $4 R_{200}$. Using these tools we address what cluster-centric distance objects lose their gas, how long it takes since crossing $4 R_{200}$ for objects to lose their gas, how host mass is linked to subhalo pre-processing, and whether some hosts are more efficient at stripping subhaloes.

This paper is organised as follows. \Sec{sec:nummethods} contains the numerical methodology used in this chapter. In \Sec{sec:nummethods-subsec:300data} we give brief reminder of the \textsc{The Three Hundred} dataset and how it was created. \Sec{sec:nummethods-subsec:sampleselect} describes how the objects in our sample and their orbital histories were selected. In \Sec{sec:nummethods-subsec:phasespace} we present the definitons and methodologies used in the analysis. Our results and discussion is presented in \Sec{sec:results}. Lastly, in \Sec{sec:conclusions} we conclude with a summary of our main findings.


\section{Numerical methods} \label{sec:nummethods}

\subsection{`The Three Hundred' Dataset} \label{sec:nummethods-subsec:300data}
 \paragraph*{The Simulations} \textsc{The Three Hundred} dataset consists of simulated clusters created by extracting 324 spherical regions of $15 h^{-1}$ Mpc radius centred on each of the most massive clusters identified at $z=0$ from the dark-matter-only MDPL2, MultiDark simulation \citep{Klypin16}\footnote{The MultiDark simulations are publicly available at \url{https://www.cosmosim.org}}. MDPL2 was simulated with a \textit{Planck} 2015 cosmology \citep{Planck15}, with $\Omega_{\rm M} = 0.307$, $\Omega_{\rm b} = 0.048$, $\Omega_\Lambda = 0.693$, $h=0.678$, $\sigma_8 = 0.823$, and $n_s=0.96$; and it contains $3840^3$ dark matter particles each of mass $1.5 \times 10^9 h^{-1}$ M$_{\odot}$ residing within a box of $1 h^{-1}$ Gpc side-length. To model the relevant baryonic physics, the 15$\hMpc$ regions extracted from the MDPL2 simulation were traced back to their initial conditions and populated with gas particles according to the Planck 15 cosmological baryonic fraction $\Omega_{\rm b}/\Omega_{\rm M} \sim 0.16$. Consequently, the resulting particles in the spherical region have a dark matter and gas particle mass resolution of $m_{\rm DM}=1.27\times 10^{9} h^{-1}$$\Msun$ and $m_{\rm gas}=2.36\times 10^{8} h^{-1}$$\Msun$, respectively. Outside the re-simulated region, to reduce the computational cost of the original MDPL2 simulation, dark matter particles are degraded with lower mass resolution particles to maintain the same large scale tidal field. The new initial conditions were simulated forward in time from $z=16.98$ to $z=0$ (in 129 saved snapshots) using \code{GADGET-X} \citep{Beck16}, using a Plummer equivalent softening of $6.5\hkpc$ for both the dark matter and baryonic component. \code{GADGET-X} is  a modified version of \code{GADGET3}  (an updated version of \code{GADGET2} by \citealt{springel_gadget2_2005}), with a modern Smooth Particle Hydrodynamics (SPH) scheme which improves the treatment of gas particles in the presence of dynamical instabilities and mixing processes, alleviates clumpiness instabilities, and reduces the viscosity away from shock regions \citep{Beck16,Sembolini16b}. Results of simulations of galaxy clusters based on \code{GADGET-X} have been presented in several previous papers \citep[e.g.][]{Rasia15,Planelles17} and in the \textit{nIFTy cluster comparison} project \citep{Sembolini16a,Elahi16,Cui16,Arthur17}.

\paragraph*{The Halo Finding} The halo analysis was done using the \code{AHF}\footnote{\url{http://popia.ft.uam.es/AHF}} halo finder \citep{Gill04a,Knollmann09}. \code{AHF} locates local overdensities in an adaptively smoothed density field as potential halo centres. Thus, it automatically creates a hierarchical structure between haloes and substructure, i.e. subhaloes, subsubhaloes, etc. The radius of a halo $R_{200}$ (and its corresponding enclosed mass $M_{200}$) are calculated as the radius $R$ at which the density $\rho(R)=M(<R)/(4\pi R^{3}/3)$ drops below $200\rho_{\rm crit}$, where $\rho_{\rm crit}$ is the critical density of the Universe at a given redshift $z$.

We maintain the definitions used in \citet{Arthur19}. We use the term \textit{halo} to refer to an object comprised of a collection of dark matter and baryonic particles, as classified by \code{AHF}, which does not reside within $R_{200}$ of another halo. Conversely, \textit{subhalo} refers to an object that does reside within $R_{200}$ of another object, be it another subhalo or halo. In each of the 324 cluster regions available in the dataset we identify a \textit{main galaxy cluster halo}, defined here as the most massive object at $z=0$.

\paragraph*{The Merger Trees} The progenitors of haloes and subhaloes are followed across the snapshots with \code{MergerTree}. Objects identified at redshift $z = 0$ are tracked backwards in time and assigned main progenitors at some previous redshift. A main progenitors is defined as the object that maximises the merit function $\mathcal{M} =N_{A\cap B}^2/(N_{A} N_{B})$, where $N_A$ and $N_B$ are the number of particles in haloes $H_A$ and $H_B$, respectively, and $N_{A\cap B}$ is the number of particles that are in both $H_A$ and $H_B$. The merger trees used in the analysis allow snapshot skipping, i.e. progenitors of (sub)haloes that are not found in the previous snapshot are still searched for in earlier snapshots, recovering an otherwise truncated branch of the merger tree \citep[see][]{Wang16}.

\subsection{Sample selection} \label{sec:nummethods-subsec:sampleselect}
To build the sample, from each of the cluster regions we selected at $z=0$ all objects within $5R_{200}$ from their respective main galaxy cluster halo, and with total halo mass $M_{200} \geq 10^{11} \hMsun$. This corresponds to a median of $\sim160$ dark matter particles.

In this work we extend upon the analysis presented in \citet{Arthur19} by using the orbital histories of the $z=0$ sample up to $z=1$, which corresponds to a look-back time of $\sim 8$ Gyr. For this analysis, we only make use of the main progenitor branch of each object in our sample. Any past objects that have merged into our sample of objects as a progenitor, and not a main progenitor, will be discarded. 

As previously mentioned, each cluster region contains a main cluster halo located in the centre of the region, and like the rest of objects in the sample, a main cluster halo is tracked back by following its main progenitor branch. If an object in the sample is beyond the radial cut or falls below the mass threshold at a certain time in its orbital history, that part of its history is ignored in the analysis. For objects that do pass these criteria, at each snapshot we describe their positions and velocities relative to the position and velocity of the main progenitor of the main cluster halo of the region at that time. Major mergers during the formation of cluster haloes pose a challenge to this \citep{Behroozi15}, therefore, following the discussion in \citet{Haggar20}, we identified and removed cluster haloes whose main progenitor's position changes by more than half their radius $R_{200}(z)$ between two consecutive snapshots $[z,z+\Delta z]$. As our analysis extends up to $z=1$, this reduces the number of resimulated regions considered in the analysis from 324 to 253. We obtain 132 427 haloes and subhaloes, with mean total mass $M_{200}=1.2\times10^{12}\hMsun$ and mean gas mass $M_{\rm{gas}}=8.4\times10^{10}\hMsun$, satisfying these constraints and constituting our analysis sample. Note that, as our analysis primarily concerns the infall region of each main galaxy cluster, defined here as their $1-4R_{200}$ region, not every object can be assigned an infall redshift, i.e. the time at which an object crossed $4R_{200}$. For analysis where properties at infall are required, we instead use 85 497 haloes and subhaloes with a designated infall redshift.

\subsection{Phase space analysis} \label{sec:nummethods-subsec:phasespace}
We follow the same definitions of phase space coordinates as in \citet{Arthur19}, which in turn follows the definitions presented in \citet{Oman13}. Here we briefly describe the procedure, we refer to \citet{Arthur19} for further details. 

For every object in our sample we construct their phase-space coordinates in 3D and in projection along line-of-sight (PROJ), where the latter is designed to mimic observational studies\footnote{Although we use the same definitions as the ones presented in \citet{Arthur19}, for clarity we changed the naming convention. Our 3D and line-of-sight projection (PROJ) phase-space coordinates correspond to the `6D' and `LOS' coordinates in \citet{Arthur19}, respectively.}. Across our analysis, we normalise relative distances by the $R_{200}$ radius of the main progenitor of the main cluster halo at the corresponding redshift value $z_{\rm{snap}}$, i.e. $R_{200}(z=z_{\rm{snap}})$; and relative velocities by the cluster velocity dispersion, $\sigma$, at the corresponding redshift value. 

The 3D coordinates are defined by using all phase-space dimensions, i.e. their 3D position $\mathbf{r}$ and velocity $\mathbf{v}$, of the object $(x_{\rm h}$, $y_{\rm h}$, $z_{\rm h}$, $v_{\rm{x,h}}$, $v_{\rm{y,h}}$, $v_{\rm{z,h}})$ with respect to the same dimensions of the cluster halo $(x_{\rm c}$, $y_{\rm c}$, $z_{\rm c}$, $v_{\rm{x,c}}$, $v_{\rm{y,c}}$, $v_{\rm{z,c}})$.
In particular, we define the 3D velocity of an object, $v_{\textrm{3D}}$, as
\begin{equation}\label{eq:v6d}
    v_{\textrm{3D}} = \textrm{sgn}\left( \mathbf{r}\cdot\mathbf{v}\right) \lvert \mathbf{v} \rvert \textrm{   ,   }
\end{equation}
where $\mathbf{r}$ and $\mathbf{v}$ are the (relative) position and velocity vectors between the cluster and the object, respectively. Note that the sign of $v_{\textrm{3D}}$ allows us to disentangle which objects are infalling into or outgoing from the cluster. Finally, to estimate the corresponding 3D velocity dispersion $\sigma_{\textrm{3D}}$ of an object, we take the root mean square of the $v_{\textrm{3D}}$ distribution of subhaloes of the main cluster halo, i.e. objects within $R_{200}$ of the main cluster halo, defined by the halo finder\footnote{We use the velocity dispersion of subhaloes in the main cluster halo as it is closer to what is measured in observations. The deviation between this estimator and the velocity dispersion of the main cluster halo is less than 10 per cent \citep[e.g][]{Gill04b, Munari13}.}.

The projected coordinates are defined by arbitrarily projecting down one axis of the simulation box, in our case along the third (z-) axis of the simulation box. We note that our results are not sensitive to the choice of the projected axis. We use the two remaining spatial coordinates, i.e $(x$, $y)$, to calculate the distance from the cluster,
\begin{equation}\label{eq:rlos}
    R_{\textrm{PROJ}} = \sqrt{ \left( x_{\rm h} - x_{\rm c} \right)^2 +  \left( y_{\rm h} - y_{\rm c} \right)^2} \textrm{   ,   }
\end{equation}
A more in-depth study of the  projected phase space of the sample (at $z=0$) is presented in \citet{Arthur19}.


\section{Results} \label{sec:results}

The analysis is split into several parts. In the first section we examine cluster build up in and beyond $R_{200}$. We then stack the orbital histories from our sample in order to construct gas-loss relations with respect to cluster-centric distance and time-since-infall. Lastly, we examine how subhalo gas disruption is linked to their host environment. Note that, to determine the gas evolution of our sample, we simply use the mass of all the gas particles within $R_{200}$ of an object.

\subsection{Cluster and infall region build up} \label{sec:results-subsec:buildup}
\begin{figure}
\includegraphics[width=\columnwidth]{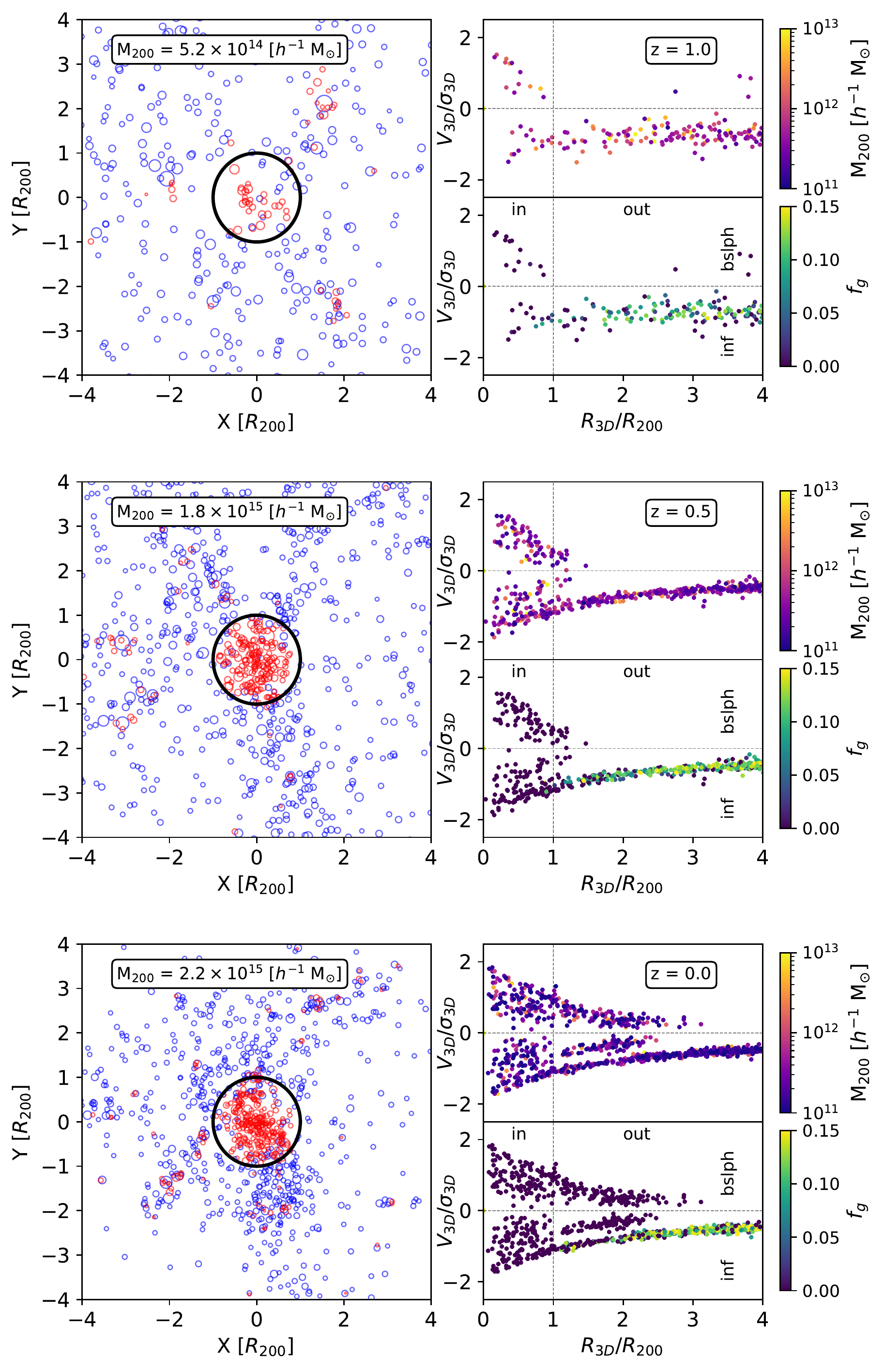}
\caption{One galaxy cluster resimulation from the sample shown at three different epochs. The top, middle and bottom panels depict the cluster at $z = 1.0$, $z = 0.5$ and $z = 0$, respectively. The halo (blue circles) and subhalo (red circles) projected distribution is shown in the left-hand panels at each epoch, and the phase-space distribution of these objects is shown in the right-hand panels. The black circle in the left-hand panels shows where $R_{200}$ of the cluster halo is. The two phase-space planes at each epoch are coloured by object mass, $M_{200}$, and object gas fraction, $f_{\rm g}$. The vertical dotted line separates the objects inside the cluster halo (`in') from the ones outside it (`out'), while the horizontal one differentiates the backsplash population (`bsplsh') from the infalling one (`inf').}
\label{fig:cluster_3pochs}
\end{figure}

In \Fig{fig:cluster_3pochs} we show the halo and subhalo spatial distribution and phase-space information at three different snapshots in time around the same cluster. The top, middle and bottom panels depict the cluster at $z = 1.0$, $z=0.5$ and $z = 0$, respectively. In each panel in the left column we represent the projected distribution of haloes with blue circles and subhaloes (be it the main galaxy cluster halo or any other halo) with red circles. The size of the circles represents the relative mass of the objects and the black circle shows the $R_{200}$ radius of the cluster halo. In the right column we present the phase-space information at each redshift coloured by their mass $M_{200}$ and their gas fraction $f_{\rm g}$, defined as the mass of all the gas particles within $R_{200}$ of an object over its total mass. The vertical dotted line separates the objects inside the cluster halo (`in') from the ones outside it (`out'), and the horizontal dotted line differentiates the backsplash population (`bsplsh') from the infalling one (`inf').

From the snapshots it is apparent that the cluster accretes more objects as time progresses. At $z = 1.0$, the infall region is sparsely populated and the cluster halo itself does not contain much substructure. However, by $z = 0.5$, the infall region has accreted considerably more substructure, with some objects apparently coalesced into filamentary structures. Alongside this, the infall region contains more host environments at $z = 0.5$ than at $z = 1.0$, potentially leading to more pre-processing at this epoch. At $z = 0$, the cluster halo still contains a considerable amount of substructure, but the infall region looks comparatively different to previous epochs. Overall, we observe that there are many more haloes and subhaloes more isotropically distributed around $R_{200}$.

By examining the phase-space planes in the right-hand panels at each epoch, we can extract more information about how the infalling objects are accreting onto the cluster. At $z = 1.0$, the infalling population has a substantial amount of scatter in velocity. This might be due to the lack of virialisation of the cluster, or due to the influence of the host environment at $\sim 2R_{200}$. By $z = 0.5$, the phase-space plane shows that the infall region contains a tight infalling branch, with little scatter in velocity. We identify a few objects that are scattered off the main branch, probably dynamic subhaloes falling into host environments. These objects are particularly good candidates to look for signs of pre-processing, especially considering they are also gas-poor. As previously seen from the gas fractions in \citet{Arthur19}, infalling objects are substantially more gas-poor at $< 1.5R_{200}$. At $z = 0.5$ the cluster contains many more subhaloes than at the earlier epoch; some of these objects are on their first infall, whilst others have already passed their pericentre passage, but have not travelled beyond $R_{200}$ to form the backsplash population. However, by $z = 0$ there is a distinct backsplash population that extends out to  $\sim 2.5R_{200}$. Interestingly, in this cluster one can see a clear group of objects that are on their second infall, located between $\sim 1-2.5R_{200}$. The extent of the backsplash and second infalling population explains the numbers of objects seen just outside $R_{200}$ in the left-hand panel at $z = 0$.

\begin{figure}
\includegraphics[width=\columnwidth]{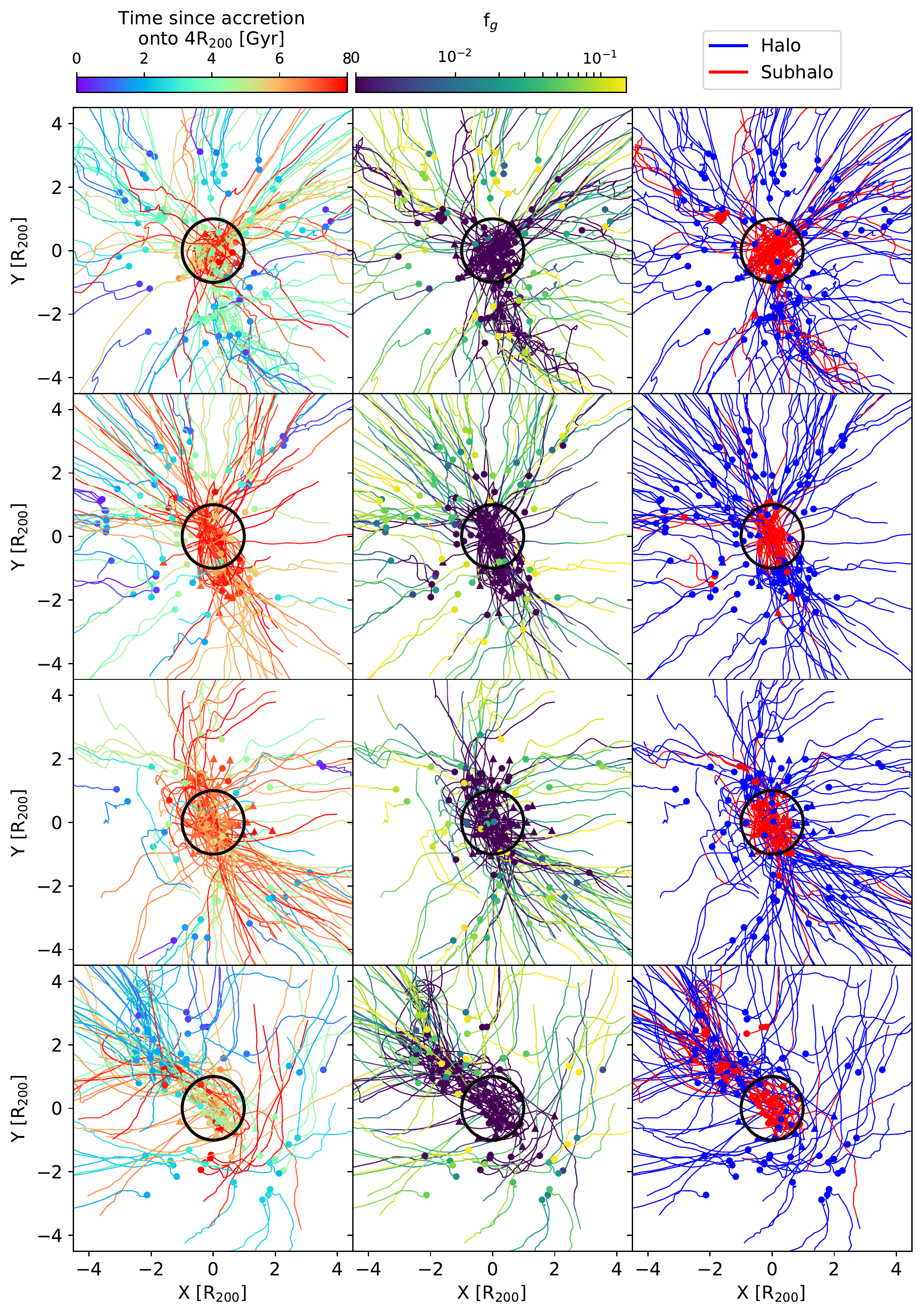}
\caption{The orbital histories of infalling objects around four example clusters are shown. Each row contains a different cluster, whilst each column colours the orbital history differently depending on a certain property. The left-hand column colours the whole orbital history by the time since an object crossed $4R_{200}$. The middle and right-hand columns contain orbital histories that have coloured sections based on the gas fraction and their subhalo status at that position, respectively. Infalling objects at $z = 0$ are marked with circle markers, backsplash objects with triangle markers. For clarity, only a subsample of 150 infalling objects are shown in each resimulation.}
\label{fig:cluster_props}
\end{figure}

To support the snapshot-only picture provided in \Fig{fig:cluster_3pochs}, we show the orbital histories of infalling objects around four example clusters in \Fig{fig:cluster_props}. For clarity, we show only the histories of a small random subsample of 150 objects in each resimulation. To understand in more detail how each cluster and infall region has been built up, the left-hand column contains the projected orbital histories of the infalling objects coloured by their time since accretion onto $4R_{200}$, i.e. since their crossing of $4R_{200}$. On the other hand, to find how the gas in each infalling object behaves as they infall, we colour the halo orbital histories by the gas fraction they have in that position. We have also coloured the orbital histories in the right-hand column by their subhalo status at the corresponding point in time, i.e. if the object resides within another halo or not at the given time. Additionally, we mark infalling objects at $z = 0$ with circle markers and backsplash ones with triangle markers. Cross-correlating this information allows us to examine in detail how our example clusters are built, and how the gaseous
properties of our infalling objects are affected by their local environment or subhalo status.

In all four example cluster regions, we identify a population of objects that have accreted early that either become part of the backsplash or virialised populations at $z = 0$. As our sample only contains objects that survive to $z = 0$, no mergers can be seen in the orbital histories. These early accreting objects are nearly always gas-poor as soon as they enter the cluster halo, and we can see how the backsplash objects can contaminate infall region observations designed for identifying pre-processing. However, as stated in \citet{DeLucia12}, as these objects have accreted early, they are older and therefore might had more time to be quenched by secular processes. From these left-hand panels, we can see the build up of each cluster is fairly diverse. The top-middle cluster presents a more in-situ build up, where many of the early accreting objects have not come in from a large distance or as part of a merger. Contrary to this, the top cluster appears to have been more built up by many objects that have traversed the whole infall region. However, we do note that our analysis is limited between $z = 0$ and $z = 1$. Some of these clusters may have undergone drastic transitions very early, which we have no knowledge about in this analysis. Nevertheless, a lookback time of  $\sim 8$ Gyr should be satisfactory to build up a good picture of each cluster’s accretion history. 

The infall region is also structured differently in each resimulation. The bottom-middle cluster appears to be funnelling in material through what appears to be filamentary structures in projection, whilst most of the material in the bottom row is being brought in as part of a large object on the top left-hand side of the panels. Despite the diversity observed in these regions, they all display signs of filamentary structure. 

Whereas the objects that have travelled into the cluster halo are nearly always gas-poor, the objects in the infalling regions are not,  although there are some caveats and these exceptions are possible pre-processing candidates. For instance, there are objects in the top three rows that become gas-poor in the infall region, before they entered the main cluster halo, and this gas disruption corresponds well with them being a subhalo of some other infalling object. Therefore, it is quite possible that these objects are being stripped as they fall into host environments at large cluster-centric radii. However, there are some haloes which become gas-poor in the infall region, which may be because of some local filamentary gas environment or some secular process. The bottom cluster shows the most obvious case of pre-processing. Many gas-poor objects on the left-hand side of the panels are clearly orbiting some large host halo, which is stripping their gas before reaching the main cluster halo.

It’s clear from \Fig{fig:cluster_props} that the accretion histories of galaxy clusters are extremely diverse, and difficult to generalise and describe by only one or two metrics. It is also evident that within our resimulations, there is some evidence for object gas disruption within both the cluster halo and the infall region. Whereas \Fig{fig:cluster_props} contains a substantial amount of detail about this, we next go on to stack all object orbital histories from each resimulation, in order to extract robust general trends and to learn more about the extent of gaseous disruption since objects crossed $4R_{200}$.

\subsection{How gas-loss since infall relates to cluster-centric distance} \label{sec:results-subsec:gasloss_distance}
\begin{figure}
\includegraphics[width=\columnwidth]{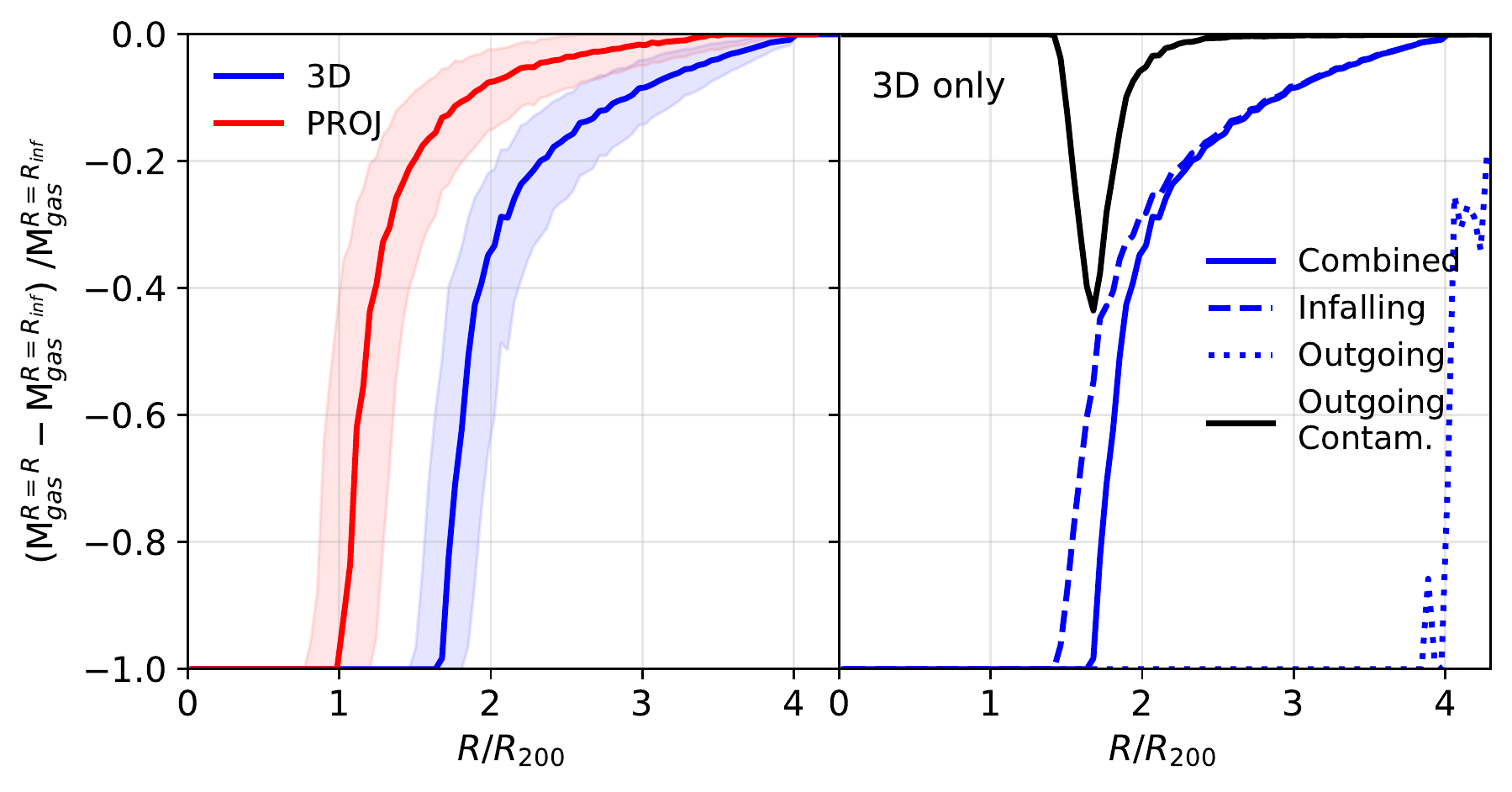}
\caption{Fractional gas-loss as a function of cluster-centric distance. See text for details about the calculation. The left-hand panel contains the median fractional gas-loss radial distribution for both the 3D and PROJ perspective in blue and red respectively, where both perspectives are constructed as in \citet{Arthur19}. The right-hand panel contains the 3D fractional gas-loss trend (solid-blue) again for comparison. The trends have been decomposed into two: one containing only infalling (dashed-blue) objects and another containing only outgoing (dotted-blue) objects. See text for details about disentangling these populations. The solid black line represents the outgoing contamination in the combined trend.}
\label{fig:deltaMgas_los_6d}
\end{figure}

In this section we stack orbital histories to assess whether, on average, there is a characteristic radius at which infalling haloes and subhaloes lose their gas since infall. This section directly follows up on the work done in \citet{Arthur19}, where it was found that the instantaneous gas fractions of infalling objects dropped to approximately zero at $\sim 1.5R_{200}$ and $ R_{200}$ in the 3D and PROJ perspectives, respectively. In this case, the temporal study actually allows us to answer whether the gas is lost at these points in space.
In \Fig{fig:deltaMgas_los_6d} we present these radial relations, where the left-hand panel contains the median fractional gas-loss as a function of cluster-centric distance in both 3D and PROJ. For every object, their gas mass is recorded at infall ($R_{\rm{inf}} = 4R_{200}$), and this is then used to calculate the fractional gas-loss at each snapshot for every orbital history. As remarked in \Sec{sec:nummethods-subsec:sampleselect}, to determine properties at infall we only use haloes and subhaloes with an assigned infall time. The radial trends are found by stacking in radial bins and then extracting median gas-loss fractions in each bin with corresponding 40 and 60 percentiles.

From the 3D distribution in the left-hand panel, infalling objects start to lose their gas gradually, beginning at $\sim 3R_{200}$ until $\sim 2R_{200}$, where objects have lost $\sim 30$ per cent of their gas on average. At this point, there is a dramatic upturn in object gas-loss, and between $\sim 2R_{200}$ and $\sim 1.7R_{200}$, infalling (sub)haloes appear to lose all of their gas. Our results here corroborate and further the results presented in \citet{Arthur19}, as it is now clear infalling objects start losing the majority of their gas close to $\sim 2R_{200}$. Previous studies \citep[e.g.][]{Behroozi14, Niemiec19} found that infalling dark matter haloes start losing their dark matter around similar cluster-centric distances, i.e. $\sim1.5-1.8R_{200}$. This characteristic radius seems to be present also for less massive haloes \citep{Buck19}. As suggested in \citet{Behroozi14}, galaxy quenching might correlate with such mass-loss processes. In the PROJ projection, the gradual loss of gas starts at $\sim 2R_{200}$, and 100 per cent gas-loss is recorded at $R_{200}$. In fact, our PROJ perspective trend agrees well with \citet{Wetzel13}, who showed that haloes containing central galaxies, are not quenched by either local environments or the cluster, until they reach $\sim 2R_{200}$. 

Like our trends, \citet{Bahe13} also found that their instantaneous radial gas fraction trends decreased with decreasing cluster-centric distance. However, they postulated that the main reason for this decline was due to contamination from
backsplash objects, which as seen in \Fig{fig:deltaMgas_los_6d}, are definitively gas-poor. We investigate this further in the right-hand panel of \Fig{fig:deltaMgas_los_6d}, which shows the radial 3D fractional gas-loss trend from the left panel, but here we have split this trend by outgoing and infalling objects. This split corresponds to a cut at $\textrm{v}_{\textrm{3D}} /\sigma_{\textrm{3D}} = 0$, where anything above and below this cut are outgoing and infalling objects, respectively. The fractional gas-loss of outgoing objects is nearly always 100 per cent, except at $\sim 4R_{200}$ , which consists of extremely dynamic, gas-rich subhaloes falling into host environments. The contamination of our trends by outgoing objects is calculated by taking the negative absolute difference between the combined and infalling trends. 

Most of the difference between these trends is seen at $\sim 1.5R_{200}$, where infalling objects appear to lose their gas $\sim 0.3R_{200}$ closer to the cluster. However, infalling objects still lose all of their gas by the characteristic $\sim 1.5R_{200}$ radius. Note that using this velocity cut does not account for second, or even third, infalling objects that have already traversed the cluster halo and been stripped. The only way to disentangle previous backsplash and $\geq 2$nd infalling objects from the first infalling population is to tag each orbital history as it passes within the cluster halo. Using a similar sample, \citet{Haggar20} showed that, depending on the dynamical state of the galaxy cluster, the median fraction of backsplash objects is $\sim40$ per cent at $2R_{200}$ for relaxed galaxy clusters, and $\sim1$ per cent for unrelaxed galaxy clusters. By $2.5R_{200}$, backsplash objects are only found around relaxed galaxy clusters. Our results show that, at similar radii, half of the gas of the infalling objects is already gone. As we do not make a distinction between relaxed and unrelaxed galaxy clusters, we expect that the contamination by these populations does not alter our trends by a substantial amount and therefore we conclude that most fractional gas-loss from infalling objects is lost on first infall \citep[see][]{Lotz19} at a characteristic radius.

\begin{figure}
\includegraphics[width=\columnwidth]{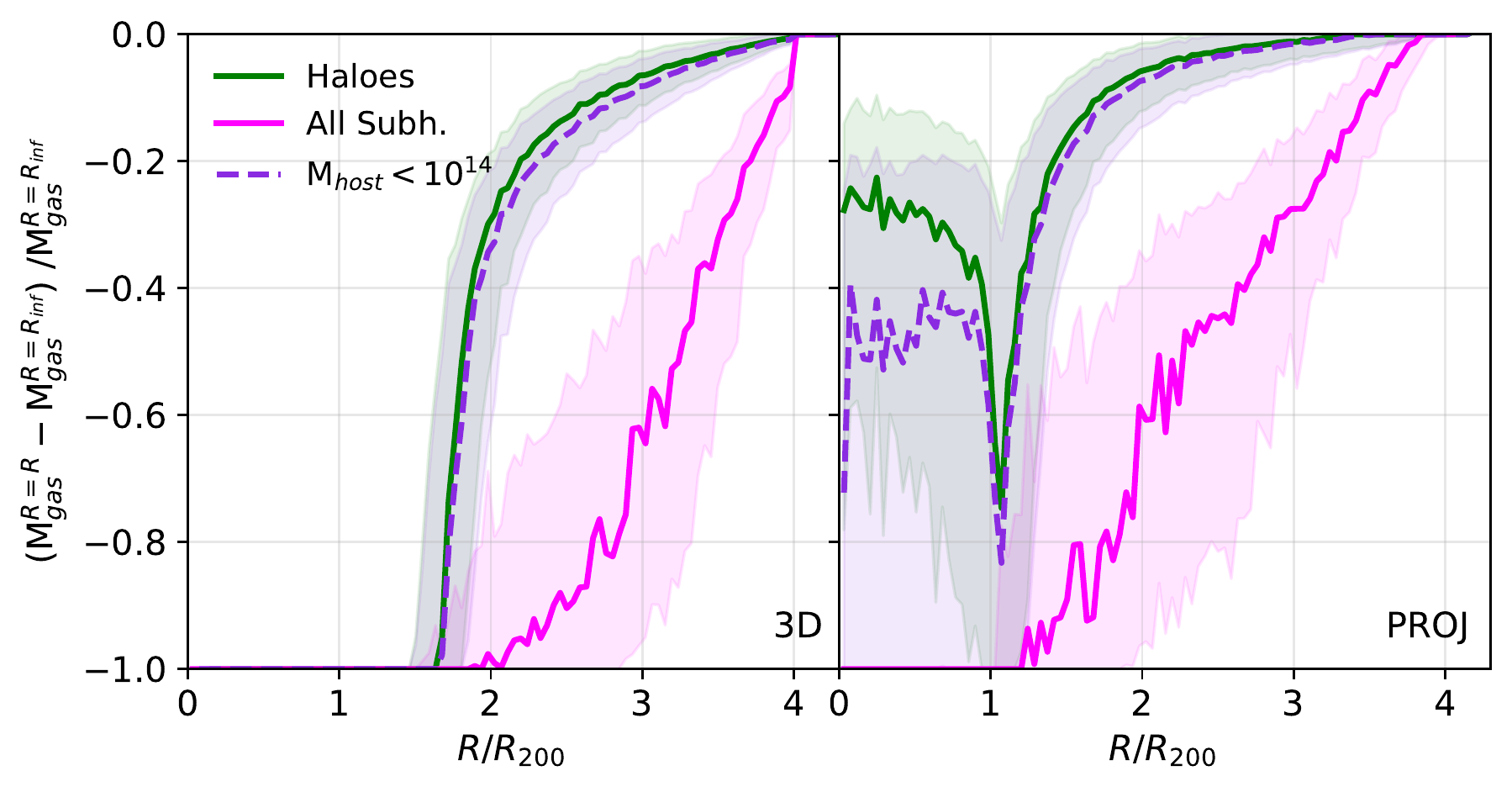}
\caption{Fractional gas-loss since infall as a function of cluster-centric distance. As in the left-hand panel from \Fig{fig:deltaMgas_los_6d}, but here the radial trends have been split by subhalo status at each radius into haloes, (all) subhaloes, and subhaloes residing in hosts with total halo mass $M_{\rm{host}}<10^{14}\hMsun$, as indicated by the legend. Here, the left- and right-hand panels show the radial trends in the 3D and PROJ perspective respectively.}
\label{fig:deltaMgas_halosubh_los_6d}
\end{figure}

In \Fig{fig:deltaMgas_halosubh_los_6d} we split the fractional gas-loss radial trends, seen in the left-hand panel in \Fig{fig:deltaMgas_los_6d}, by the subhalo status of the objects at that radius. From the 3D projection (left-hand panel), we see that subhaloes lose their gas much quicker than haloes since crossing $4R_{200}$. The halo trend still follows the combined trend in \Fig{fig:deltaMgas_los_6d} very well, as the majority of the objects in the infall region will be haloes, which also explains why there is more scatter in the subhalo-only trend. However, even with the percentiles taken into account, subhaloes have already lost $60$ per cent of their gas by $\sim 3R_{200}$, on average, whereas haloes reach the same fractional gas-loss at $\sim 1.7R_{200}$. While this is compelling evidence for pre-processing of subhaloes by host environments, we further separate the subhaloes in the sample by their host halo mass. We note that the radial cut used to determine the sample region at $z=0$, i.e. $5R_{200}$ from the most massive object, contains a diverse range of haloes and subhaloes masses, including objects with total mass $M_{200}>10^{14}\hMsun$. In order to differentiate galaxy disruption due to sub-dominant environments (i.e. pre-processing) from disruption exerted by cluster-mass hosts infalling onto the main galaxy cluster (i.e. cluster quenching), we remove from the subhalo sample the subhaloes residing in cluster-mass host haloes. The fractional gas-loss of the subhaloes in group-mass and low-mass hosts is shown as a dashed violet line. We can clearly see that the fractional gas-loss of these subhaloes follows essentially the same radial disruption trend observed in the halo population in 3D. For the PROJ projection, haloes, and subhaloes in group-mass and low-mass hosts follow a similar trend up to $R_{200}$, from which the subhaloes have $\sim 20$ per cent less gas than haloes, albeit with more scatter. From these relations we conclude that subhaloes that reside within cluster-mass host haloes during their infall onto the main cluster halo, i.e. subhaloes of cluster-mass host haloes found in the infall region of the main cluster halo, are quickly depleted of their gas and dominate the radial gas-loss trend if we consider all subhaloes regardless of the mass of their host haloes. 

Some recent studies \citep{vanDenBosch18a,vanDenBosch18b} have found that subhaloes in hydrodynamic simulations are prone to enhanced artificial tidal stripping due to inadequate mass or spatial resolution, even at a nominal ‘resolved’ mass cut (e.g. $10^{11} \hMsun$ , which is used in this analysis). For example, if the intrahalo medium is not well resolved gas particles become easily unbound. As a consequence, the gas-loss (or gain) trend found for infalling objects simply translates the motion towards higher or lower density regions. In light of these issues, we consider the characteristic radius $1.7R_{200}$ ($1.5R_{200}$ when removing contamination from outgoing objects) below which infalling objects lose their gas as an upper-bound of the accretion shock radius. Nevertheless, it is most probable that the fractional gas-loss trends are a combination of environmental stripping by host environments, secular processes, and enhanced numerical stripping. The interesting question here is how dominant the environmental effect is, and will be an area of future work.

We note here that our trends line up well with those seen in \citet{Wetzel13}, but this time we have identified that the halo population, which contain central galaxies, seem to be affected by the cluster at $\sim 2R_{200}$. This is especially true in the PROJ perspective, whereas in the 3D perspective, haloes seem to be affected somewhat further out. We note that any differences in gas-loss as a function of cluster-centric distance between the halo and subhalo populations are somewhat washed out by projection effects intrinsic to the PROJ perspective.

\subsection{How gas-loss relates to time-since-infall} \label{sec:results-subsec:gasloss_inftime}
\begin{figure*}
\includegraphics[width=0.8\textwidth]{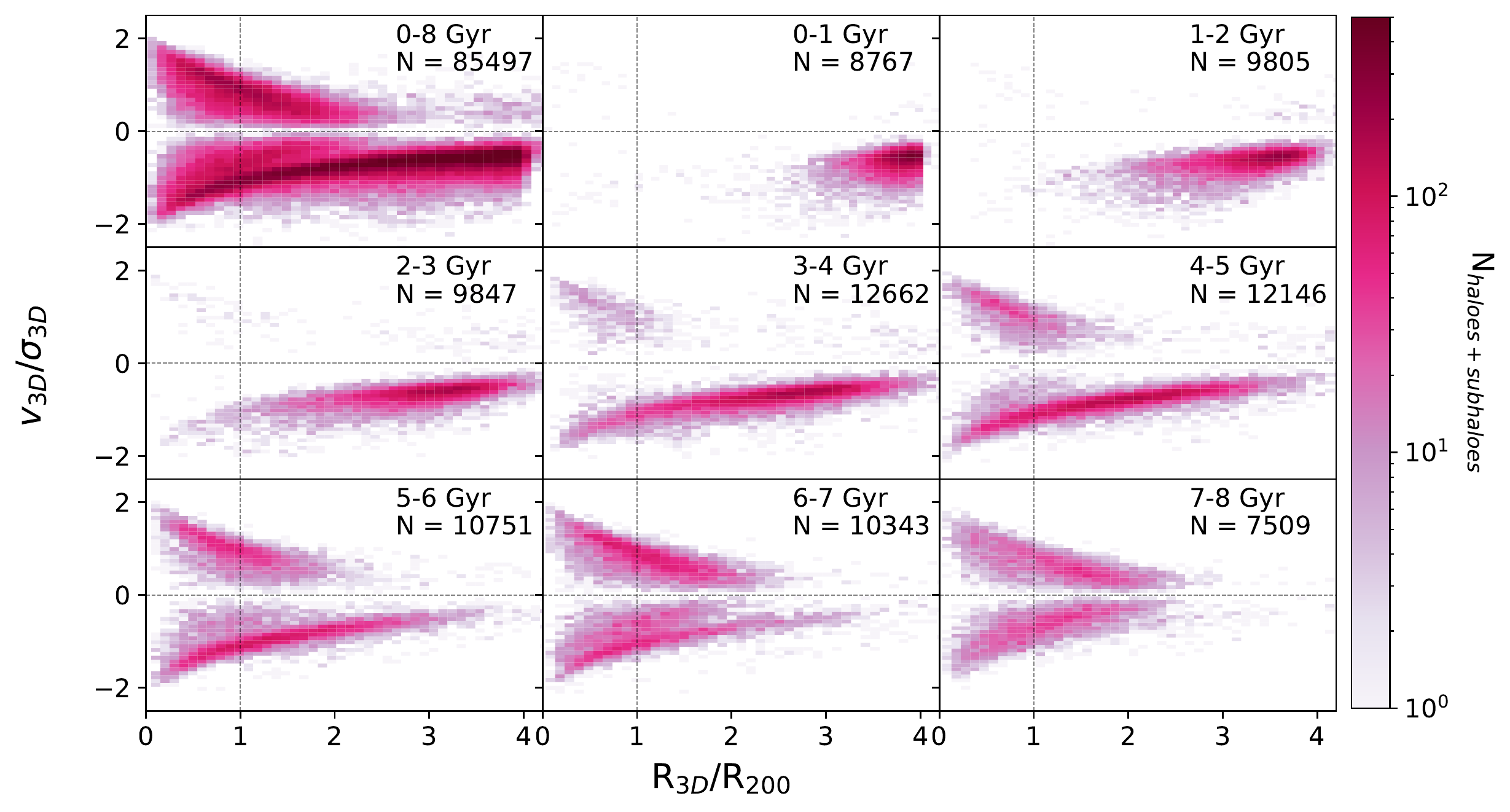}
\caption{The stacked phase-space plane at $z=0$ in the 3D perspective, as a function of time-since-infall. The following panels are found by splitting the top-left panel in time-since-infall, as indicated in the top-right part of each panel. Each panel is coloured by the number of objects in each bin,  which is given by the colour bar.}
\label{fig:phasespace_nhalos_time}
\end{figure*}

In this section we use the orbital histories of our sample to investigate how gas-loss since crossing $4R_{200}$ relates to time since crossing that radius. \citet{Bahe15} first used the \textsc{GIMIC} simulations to characterise how satellites were quenched as a function of time since passing $R_{200}$. We extend their analysis by examining, on average, how the infall region builds up as a function of time-since-infall and how an object's gaseous properties are also affected. In order to acquire a clear picture of how \textsc{The Three Hundred} clusters generally build up, we repeat the analysis done in \citet{Oman13}, which is presented in \Fig{fig:phasespace_nhalos_time}. In the top-left panel we show the stacked 3D phase-space plane from all the objects in our sample at $z = 0$, which extend the analysis done in \citet{Arthur19}. We use the orbital histories to split the objects in this plane by their time-since-infall into the following eight panels. In each panel we show the number of objects in each bin.

The top-left panel contains a distinct infalling branch, a backsplash population and a virialised population. Unsurprisingly, the objects that have recently undergone infall (i.e. $<2$ Gyr) are mostly located on the infalling branch at high cluster-centric radii. There are a small number of objects that already reside within the cluster in these panels, which are suspected to be spurious effects resulting from \code{MergerTree} mislabelling the cluster main progenitor branch. In the next three panels, that show objects that underwent infall 2-5 Gyr ago, we see that the infalling branch is well established. By 2-3 Gyr, objects are starting to accrete onto $R_{200}$ of the cluster and by 4-5 Gyr, the backsplash population is well defined. In the final three panels, the infalling branch starts to diminish at high cluster-centric radii. By 6-7 Gyr, it seems as though nearly all objects have
made their first passage, and any objects infalling at this stage are on their second, or above, re-entry. The virialised parts of these planes are well established. For the most part, our results are consistent with those in \citet{Oman13}. However, we do note that the infalling branch around the clusters in the sample takes longer to disappear; the branch in \citet{Oman13} is gone by 4-5 Gyr, but for our objects it becomes indistinct at 6-7 Gyr. Alongside this, we see backsplash objects reach much higher cluster-centric distances than those in \citet{Oman13} \citep[a detailed analysis of the backsplash population of a similar sample to the one used here can be found in][]{Haggar20}. We note that the host masses in \citet{Oman13} have a mass $> 10^{14} \hMsun$, which are comparable to our sample. However, the analysis done in \citet{Oman13} uses a different definition of halo radius, namely $R_{360b}$, the radius at which the density drops below $360\rho_{\rm bg}$, where $\rho_{\rm bg}=\Omega_{\rm M}\rho_{\rm crit}$ is the background matter density of the Universe at a given redshift $z$, and $\Omega_{\rm M}$ the matter density parameter. This difference in the volume definition (i.e. $R_{\rm 200c} \sim 0.8R_{\rm 360b}$) might explain the difference between the lifetime of the infalling branch of our sample compared to the one found in \citet{Oman13}.

\begin{figure*}
\includegraphics[width=0.8\textwidth]{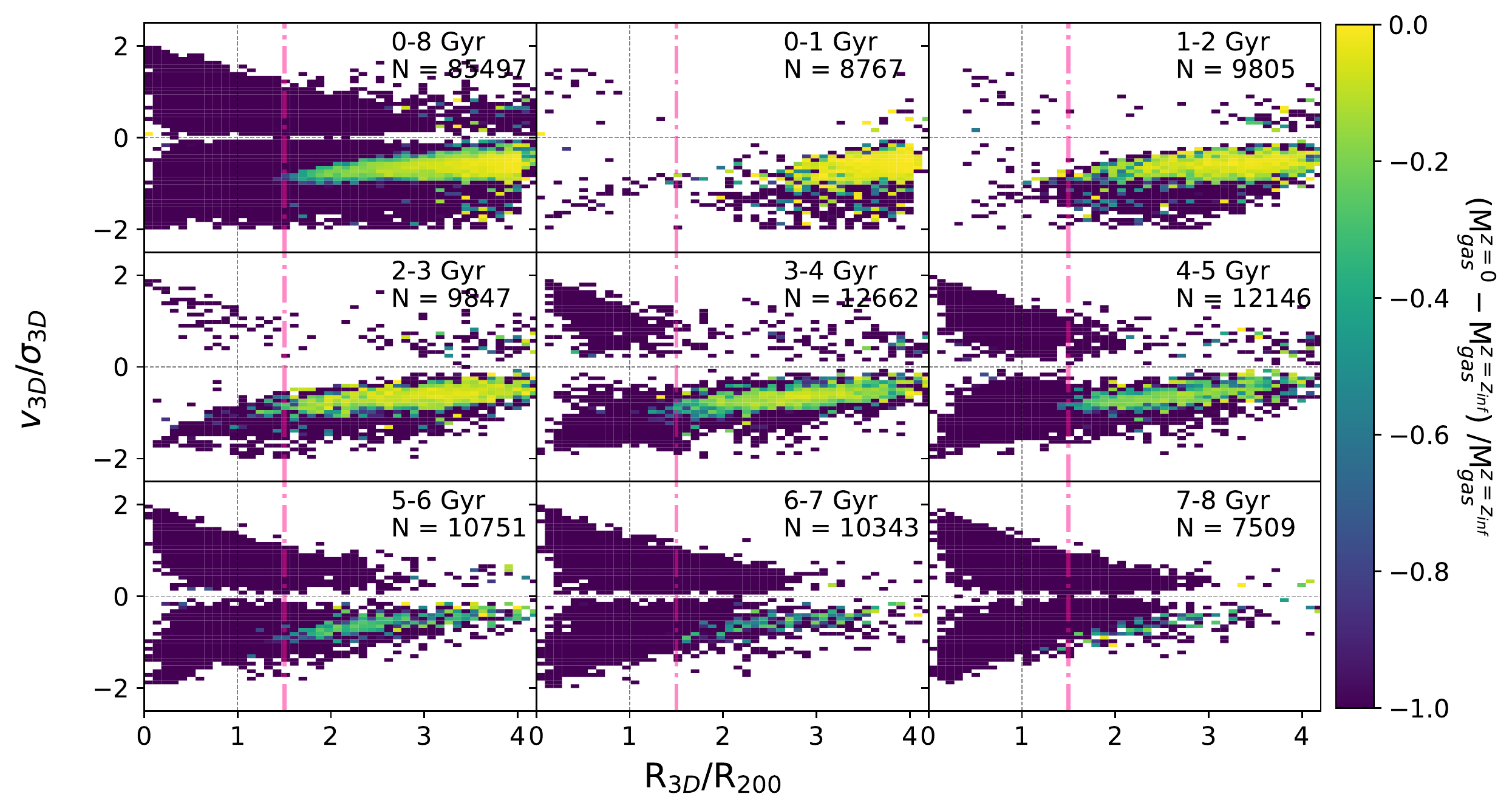}
\caption{As in \Fig{fig:phasespace_nhalos_time}, but each bin is coloured by the median fraction gas-loss since infall. The vertical pink dash-dotted line marks the $1.5R_{200}$ radius. }
\label{fig:phasespace_deltaMgas_time}
\end{figure*}

To study how an object's gas-loss since infall is related to the time-since-infall, we present in  \Fig{fig:phasespace_deltaMgas_time} the same distribution as in \Fig{fig:phasespace_nhalos_time}, but now coloured by the median fractional gas-loss in each bin. \Fig{fig:phasespace_deltaMgas_time} shows that objects, on average, lose nearly all of their gas at $\sim 1.5R_{200}$ (marked with a vertical pink dash-dotted line), which agrees with the instantaneous gas fractions presented in \citet{Arthur19}. In this case though, as we now have the orbital histories of each object at our disposal, we can confirm that nearly all infalling objects, by the time they reach $\sim 1.5$ since they entered the infall region, lose all their gas. 

In the top-middle and top-right panels, recent infalling objects have lost little gas at high cluster-centric radii, with the exception of the very dynamic pre-processing candidates. In the following three panels though, where the infalling and backsplash branches are more distinct, objects seem to have lost more of their gas. As soon as $2-4$ Gyr since infall, objects have reached the characteristic radius where the majority of gas-loss is encountered: $\sim 1.5 - 2R_{200}$. In these panels the backsplash population starts to become more distinct, but one can see that the objects forming this region have lost all of their gas. In the final three panels, which shows the objects with the latest infall, nearly all objects have lost all of their gas, with the exception of a small sample of objects that are still on their first infall. We also identify a correlation between the velocity of the objects and their gas-loss fraction: as best seen in the phase-space distribution of the objects with a time-since-infall within 3 Gyrs (top-middle, top-right, and middle-left panels), the faster they infall towards the main cluster halo, the higher their gas-loss fraction, likely due to the higher ram-pressure these objects experience as their infall velocity increases.

From \Fig{fig:phasespace_deltaMgas_time} we find that objects within the $\sim 1.5R_{200}$ region have always lost $\sim 100$ per cent of their gas, on average, regardless of their time-since-infall. This suggests that gaseous disruption in this region is mostly environmental and not driven by secular processes. Moreover, we also find that an object's gas-loss in the infall region is driven by its time-since-infall; the longer an object spends in the infall region, the more fractional gas-loss it will undergo. One plausible explanation for this is that more intermediary environments, such as groups and filaments, need longer to strip an object's gas and therefore a trend with time is more easily seen in the infall region. However, a higher time-since-infall will presumably correspond to an older object, which is more likely to be affected by secular processes in that region.

\begin{figure*}
\includegraphics[width=0.8\textwidth]{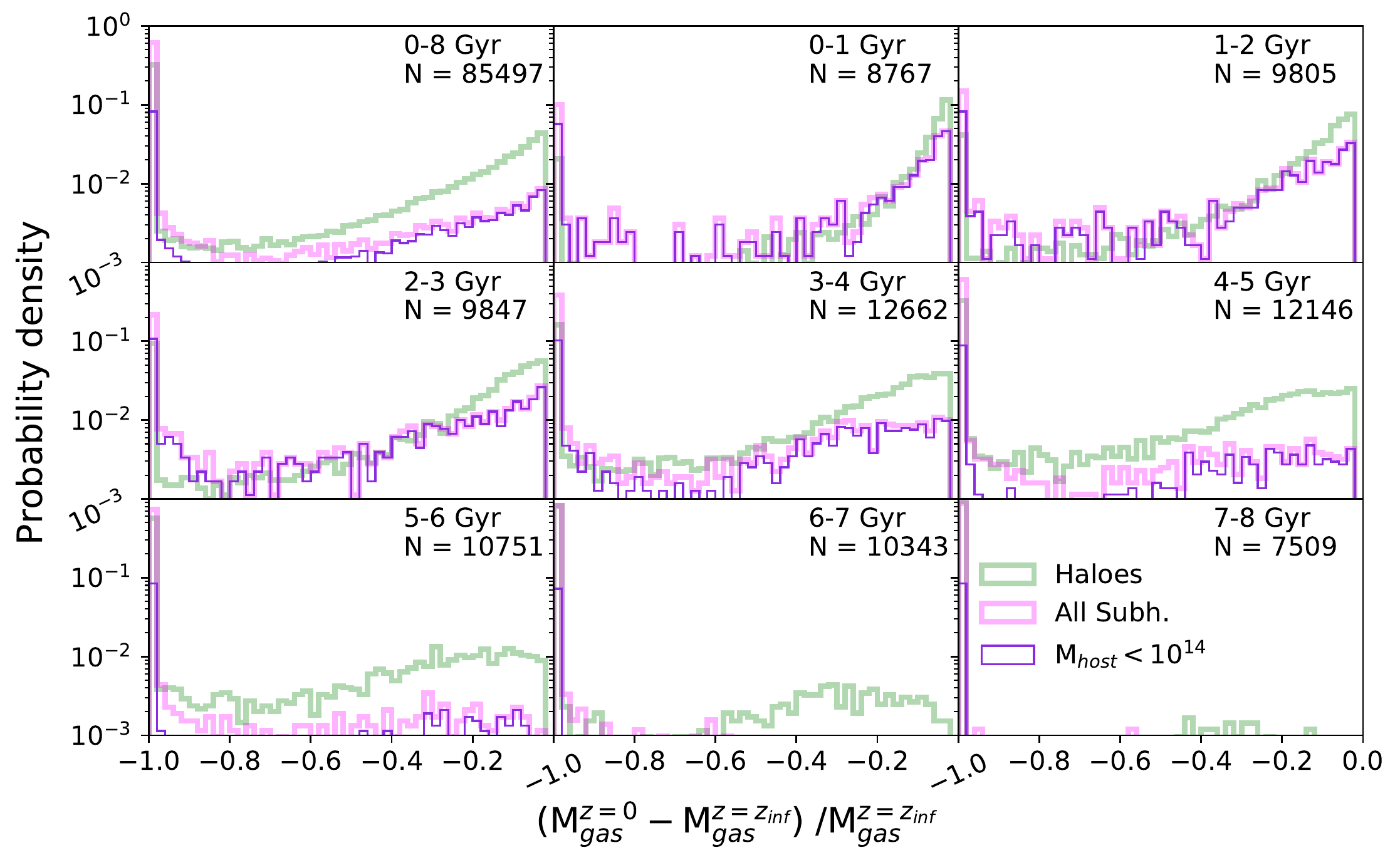}
\caption{Fractional gas-loss probability densities for all objects at $z=0$, split by haloes (green solid lines), subhaloes (magenta solid lines), and subhaloes residing in hosts with total halo mass $M_{\rm{host}}<10^{14}\hMsun$ (violet thin lines). As in \Fig{fig:phasespace_nhalos_time}, the top-left panel, which contains the whole population, is split by time-since-infall into the following panels.}
\label{fig:deltaMgas_hist_time}
\end{figure*}

To examine how haloes and subhaloes lose their gas differently since crossing $4 R_{200}$, we use the fractional gas-loss information for the population of objects at $z = 0$ to construct separate probability densities of fractional gas-loss as a function of time-since-infall. This is shown in \Fig{fig:deltaMgas_hist_time}.

As in \Fig{fig:phasespace_deltaMgas_time}, the top-left panel in \Fig{fig:deltaMgas_hist_time} contains the fractional gas-loss information for the whole population at $z = 0$. Following the discussion for \Fig{fig:deltaMgas_halosubh_los_6d}, we also show the contribution of the subhalo population residing in group-mass and low-mass host haloes to the whole subhalo sample distribution (thin violet lines)\footnote{We note that this distribution is technically not a probability density as its integral over the fractional gas-loss is the fraction of the whole subhalo sample that resides in hosts with $M_{\rm{host}}<10^{14}\hMsun$.}. In this panel, the distributions are bimodal. On average subhaloes are more likely to lose their gas than haloes, although there is a population of subhaloes which have retained their gas in this panel. The gas-poor subhaloes are, for the most part, subhaloes of the main cluster halo, though some may be objects at high cluster-centric radii that have already been pre-processed or stripped of their gas by cluster-mass host haloes. Comparing the distribution of all subhaloes with the one from sub-dominant environments we see that most subhaloes in cluster-mass hosts lost between 60 and 90 per cent of their gas since they crossed $4R_{200}$. On average, haloes retain their gas much more successfully than subhaloes. However, there is a significant population of gas-poor haloes, which potentially consist of a combination of backsplash objects or objects that have been pre-processed by some local gas environment, such as an accretion shock between $\sim 1.5 - 2R_{200}$. When we split the top-left panel by the time-since-infall, the differences between the halo and subhalo distributions become even starker. Haloes that have spent $0-1$ Gyr within the infall region are generally gas-rich and have retained most of their gas; many have lost less than $\sim 20$ per cent of their initial gas at infall. In contrast, the subhalo distributions are much more gas-depleted in this panel, even when considering subhaloes in sub-dominant environments.  From \Fig{fig:deltaMgas_hist_time} and
\Fig{fig:deltaMgas_halosubh_los_6d}, if we consider all the subhaloes in the sample, they lose their gas faster and further out from the cluster than haloes. Otherwise, they follow similar radial gas-loss trends. Interestingly, in the following panels, which show the objects with earlier infall times (i.e. that spent more time in the infall region), subhaloes lose their gas as a function of time-since-infall, but they are not fully depleted until $6-7$ Gyr, which is almost as long as the halo population. In fact, by $7-8$ Gyr, nearly all haloes and the whole subhalo population have lost all of their gas. 

\subsection{How subhalo gas-loss relates to both cluster-centric distance and time spent in the host environment} \label{sec:results-subsec:gasloss_distanceANDhost}
\begin{figure}
\includegraphics[width=\columnwidth]{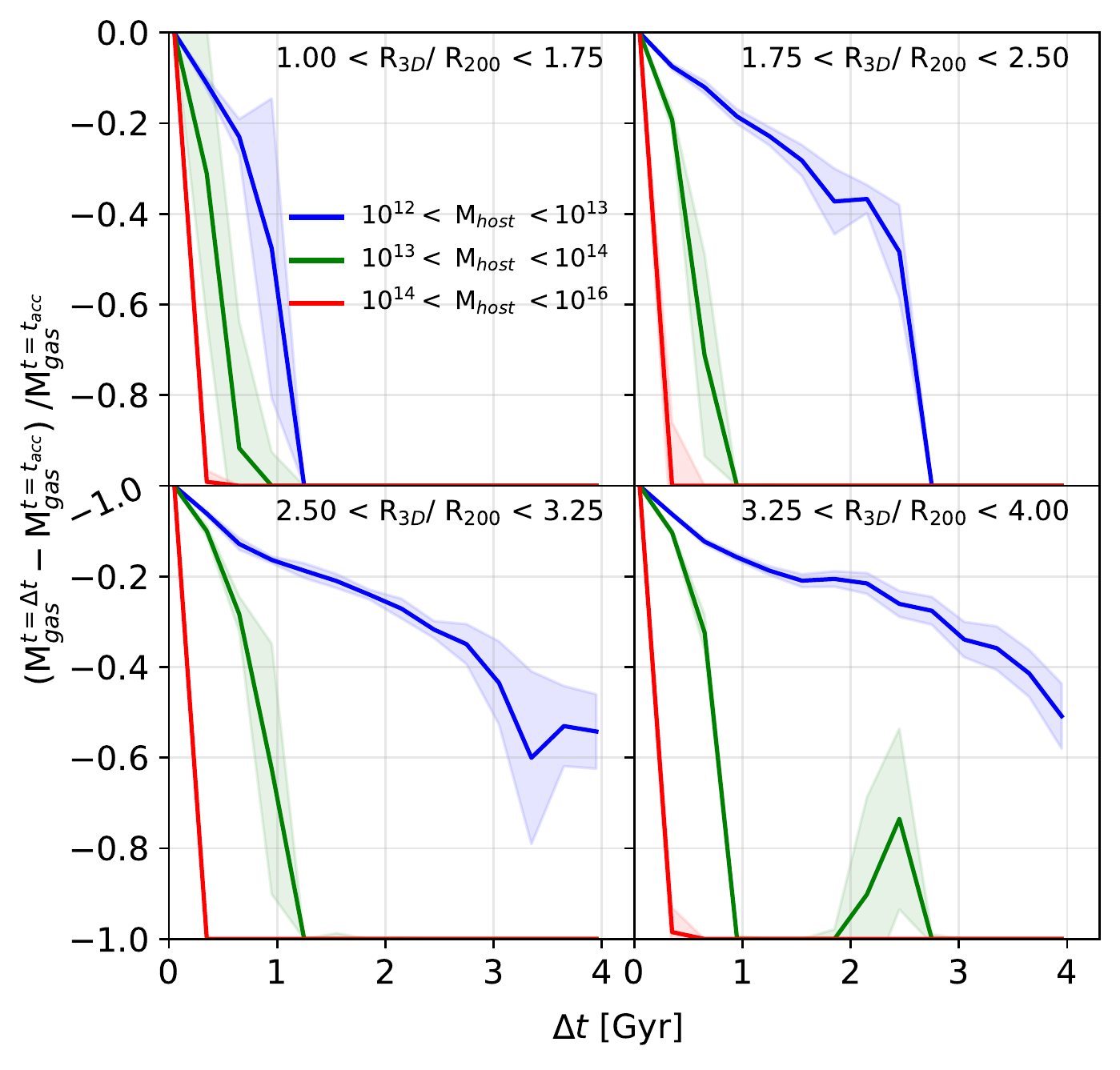}
\caption{ Subhalo fractional gas-loss as a function of time spent in host environment. The trends have been divided by mean host mass over that time, as indicated by the legend, and further subdivided by mean cluster-centric distance over that time, as indicated by each panel. Error bars are $1\sigma$ errors on the median in each time bin, calculated by bootstrap resampling.}
\label{fig:deltaMgas_rbins_time}
\end{figure}

Subhaloes generally lose more gas further from the cluster and sooner than haloes would. As seen in \Fig{fig:deltaMgas_halosubh_los_6d}, their radial trend is dominated by subhaloes in cluster-mass host haloes. Once separated from the full population, subhaloes in sub-dominant environments (e.g. group-mass haloes or filaments) follow  similar radial gas-loss relations as haloes. Barring any artificial effects, this suggests that subhaloes are being stripped off their gas before they reach $R_{200}$. In this section we examine this closely by investigating how subhalo fractional baryonic mass is lost as a function of time spent in different host haloes. By analysing the orbital histories of our objects and only considering the times at which they became subhaloes of haloes in the infall region, we can classify those parts of their histories by the mass of their host haloes, their cluster-centric radius at the time, and the time they spent in the host environment\footnote{Note that the time spent in a host halo environment differs from the time-since-infall of an object, which exclusively refers to the crossing of the 4$R_{200}$ radius of the \textit{main cluster halo}.}. This is first presented in \Fig{fig:deltaMgas_rbins_time}, which shows how subhalo gas-loss correlates with the aforementioned properties. In order to construct these trends, we use the orbital histories of the objects in the sample to record when an object accretes onto another object in the infall region to become a subhalo, and its initial gas mass at that time. The evolution of the subhalo gas content is tracked from that point onwards to construct temporal trends. Due to the fact that host haloes will accumulate mass over time, we use the mean host halo mass during the tracking time window. Likewise, this is done for cluster-centric distance\footnote{It is worth mentioning that using averaged quantities might introduce a bias in the analysis. For instance, subhaloes assigned to a radial bin might spend half of their time within their host environment in one radial bin, and half in the neighbouring one. We analysed this potential issue and found that the vast majority of subhaloes remain within their assigned radial bins for each point of the trends, with a dispersion from the mean normalised cluster-centric distance of $\sim0.2-0.3$.}. In order to control for subhalo mass, subhaloes are only tracked in host halo environments if they have an initial mass at the time they became subhaloes that is $10^{11} \hMsun \leq$ $M_{200} \leq 10^{12} \hMsun$. The shaded regions show the $1\sigma$ errors on the median in each time bin, calculated by boostrap re-sampling. 

Analysing each radial bin, we find that subhaloes lose their gas more quickly within higher mass host haloes. In the highest cluster-centric distance bin (bottom-right), subhaloes in group-mass host haloes (green) essentially lose all of their gas by $\sim 1$ Gyr, but for subhaloes residing in the lowest host mass bin (blue) it takes $\sim 4$ Gyr to lose half their initial gas.  Although we only examine the gas here, this is consistent with work done in \citet{Wetzel12, Bahe15, Roberts17} and \citet{Cora18}, who suggest that quenching and pre-processing of satellites is more dominant in higher mass hosts. This result is also compatible with the results presented in \citet{Cora19}, who showed that quenching mechanisms are less efficient in galaxies accreted by lower mass haloes, both prior and after their first infall, due to the combined effect of the milder environmental disruption exerted by low-mass host haloes, and the reduced secular quenching in low-mass subhaloes that naturally accrete onto low-mass host haloes. However, in contrast, when we consider where subhaloes lose $50$ per cent of their gas in the first radial bin, the trends only differ by less than a Gyr. This latter point is more consistent with \citet{Wetzel13}, who suggest that host mass does not influence the quenching time-scales of satellite galaxies, although their analysis was constrained to the virial radius of the clusters. Both  \citet{Wetzel13} and \citet{Cora19} observed a delay in satellite quenching after accretion onto a host, and it is possible that this delay is consistent with our subhaloes in lower mass hosts retaining some cold interstellar medium component of their gas. As for trends with host mass, looking at the lowest mass hosts (blue) of  \Fig{fig:deltaMgas_rbins_time} we see a clear radial trend. Subhaloes lose their gas quicker in these hosts the closer they are to the main cluster halo. Subhaloes in these hosts in the furthest bin lose $40$ per cent of their gas by $\sim 4$ Gyr, but in the closest bin they lose all of their gas by $\sim 1.5$ Gyr. These radial differences are mainly seen in the last 80 per cent of gas, as across all bins subhaloes in the lowest mass hosts reach $\sim 20$ per cent gas-loss by $\sim 1$ Gyr. These results suggest that whilst the low mass hosts are efficient in stripping the first $\sim 20$ per cent of gas, being in closer proximity of the main cluster halo is enough for the main cluster environment to take over and strip the remaining $\sim 80$ per cent. Interestingly, no radial trends are found for group-mass or cluster-mass hosts; they are just as efficient at removing all gas from subhaloes at any cluster-centric radii. This is good evidence for gas disruption of subhaloes by group environments (green-line) at high cluster-centric distances.

\begin{figure}
\includegraphics[width=\columnwidth]{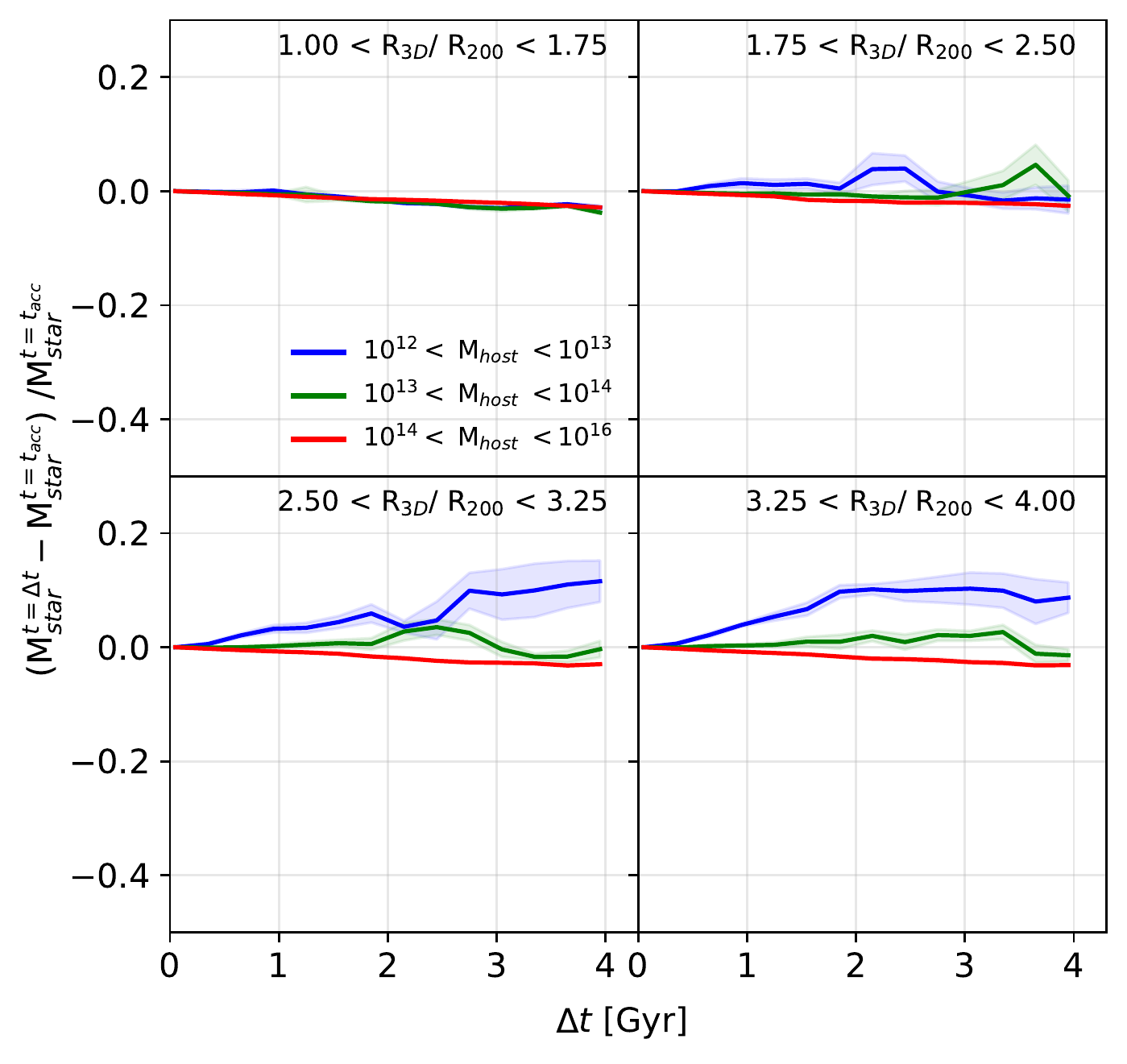}
\caption{  As in \Fig{fig:deltaMgas_rbins_time}, but shows the fractional stellar loss of subhaloes as a
function of time spent in a host environment.}
\label{fig:deltaMstar_rbins_time}
\end{figure}

By only examining the fractional gas-loss of subhaloes, it is difficult to disentangle whether any lost gas is being stripped (external processes), ejected (internal processes), or whether it is being turned into stars over time. In light of this we next present \Fig{fig:deltaMstar_rbins_time}, which depicts the same information as in \Fig{fig:deltaMgas_rbins_time}, but instead shows the fractional stellar loss of subhaloes as they fall into hosts haloes.

Firstly, we identify a radial trend in the stellar mass loss of subhaloes within low-mass hosts (blue): the stellar mass loss increases as we move closer to the main cluster halo. For the other two subhalo populations, i.e. subhaloes of group-mass (green) and of cluster-mass (red) hosts, the stellar mass loss radial trends are fairly similar, with the exception of the innermost radial bin (top-left panel), in which all three subhalo populations show a light correlation between their stellar mass loss and the time spent in their respective host environments, regardless of the mass of their host halo. On the other hand, subhaloes residing in low-mass and group-mass hosts display a gain of stellar mass outside the innermost radial bin. In the highest radial bin (bottom-right panel), there is a $\sim 10$ per cent difference in the stellar mass change between the subhalo population of the lowest and highest host masses after 4 Gyrs inside the host environment. Nevertheless, for high-mass hosts (red), the trends in each radial bin show no increase in stellar material, and in the stacked data it seems as though the stellar material only decreases as a function of time spent in the host environment. These results are in partial disagreement with \citet{Wetzel12} and \citet{Wetzel13}, who suggest that satellites are still allowed to form stars and grow in mass before a delayed, but rapid quenching. In our results we do not find star formation for subhaloes residing in cluster-mass hosts. For massive galaxies, however, \citet{Cora19} showed that the duration of the phase following the delay might, in fact, be of the same order of magnitude as the delay phase. 

It is possible that the star-loss found in such subhaloes, as opposed to the stellar gain observed in the other populations, is not solely a consequence of a lack of star formation. Such subhaloes might indeed form stars but stripping or ejection of stellar material from environmental or secular processes might outweigh the star formation taking place in the subhaloes, resulting in a net loss of both gas and stellar material. We concede that the number of star particles is less than statistically significant for some subhaloes in the sample. As a first approximation, taking into account the cosmic baryon fraction of our simulations, a population of subhaloes with mass $10^{11} \hMsun \leq$ $M_{200} \leq 10^{12} \hMsun$ will contain between $\sim 70$ to $\sim 700$ star particles. For subhaloes with only few star particles, one star particle being stripped from a subhalo could equate to as much as a $\sim 10$ fractional star-loss, but this is presumably smoothed out by using the stacked median in temporal bins. However, even considering this, we do find that, on average, the stellar mass of subhaloes in massive hosts is not increasing as a function of time spent in the host. Therefore, we may statistically infer that subhalo gas is really being lost by either stripping or ejection in such subhaloes, while for subhaloes residing in group and low-mass hosts, gas-loss is boosted by the star mass gain identified in \Fig{fig:deltaMstar_rbins_time}. In order to differentiate between these processes and estimate their relative importance, a thorough tracking analysis is needed that examines the history of each and every gas and star particle situated in infalling subhaloes.

\begin{figure}
\includegraphics[width=\columnwidth]{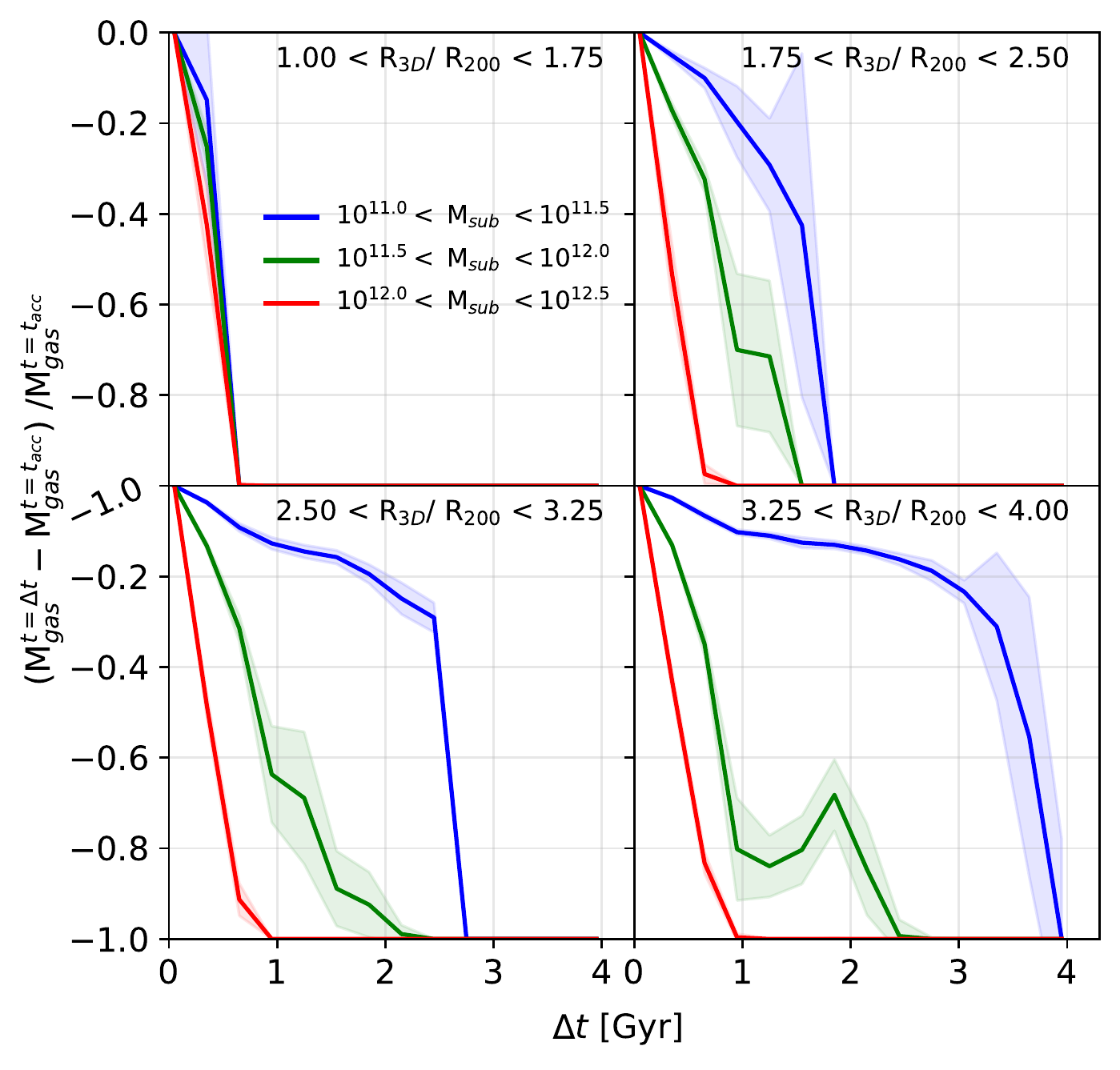}
\caption{ As in \Fig{fig:deltaMgas_rbins_time}, but the different lines refer to the mass each subhalo had on
accretion into the host environment, as indicated by the line colour and legend.}
\label{fig:deltaMgas_subh_rbins_time}
\end{figure}

While we suggest that it is compelling that gas disruption by host environments, either by sub-dominant environments or by cluster-mass hosts, is the main cause for enhanced subhalo gas-loss during their infall onto the main galaxy cluster, it is most likely a mixture of actual gas disruption, secular processes and enhanced numerical stripping of subhaloes due to inadequate resolution \citep{vanDenBosch18a,vanDenBosch18b}. To conclude our analysis, we finish by investigating how subhalo gas-loss is related to number of particles in our subhaloes. In \Fig{fig:deltaMgas_subh_rbins_time} we present the fractional gas-loss of subhaloes, as in \Fig{fig:deltaMgas_rbins_time}, but here the trends are split by the initial mass each subhalo has when it accretes onto the host. 

We note that for this part of the analysis we do not make the same initial subhalo mass cuts that we used in \Fig{fig:deltaMgas_rbins_time} and \Fig{fig:deltaMstar_rbins_time}. What is striking on first glance is that we find that group and low-mass subhaloes retain their gas for longer in all radial bins, with the exception of the first radial bin (top-left) in which there is no apparent difference between the populations. Analysing the host mass of these subhalo populations we confirm that larger mass subhaloes reside in larger mass host haloes, which as seen from \Fig{fig:deltaMgas_rbins_time}, are more efficient environments for subhaloes to lose their gas in. For the innermost radial bin, similar to what we found \Fig{fig:deltaMgas_rbins_time}, they all lose their gas at similar times. In light of this result, we now are able to better understand the radial correlation seen in the stellar mass change in subhaloes of group and low-mass hosts and their cluster-centric distance in \Fig{fig:deltaMstar_rbins_time}. One possible scenario is that, as subhaloes within more massive hosts would accrete with larger velocities in the high density environment of their host, the ram pressure they would experience would be greater than in comparable subhaloes in lower mass hosts. As discussed in \citet{Cora19}, secular processes (e.g. feedback from active galactic nuclei) play a major role in the decline of the star formation for high-mass galaxies both before and after their infall time, which they defined as the time at which a galaxy becomes a satellite for the first time. As subhaloes in group and low-mass hosts tend to be on the lower end of the subhalo mass distribution, they are able to form stars even after 4 Gyrs since they entered their host halo environment. 

These trends lend support for these subhaloes losing gas by physical processes, rather than any artificially enhanced stripping due to the number of particles used to describe such subhaloes, as otherwise one would expect these artificial effects to lead to enhanced stripping in the lower mass subhaloes (as we discussed in \Sec{sec:results-subsec:gasloss_distance}). Presumably the host environment dominates this numerical effect, thus more massive subhaloes lose their gas quicker, and lower mass subhaloes retain their gas for longer, despite having less dark matter particles. Although these trends diminish the numerically enhanced stripping argument, they do not allow us to differentiate on whether subhaloes are losing their gas via stripping or secular processes, as in larger mass subhaloes one would expect more extreme feedback schemes to be implemented in order to control stellar production.


\section{Conclusions} \label{sec:conclusions}

We used haloes from \textsc{The Three Hundred} project, a suite of 324 zoomed galaxy cluster simulations that extend $> 5R_{200}$ of the main cluster halo, to investigate the level of gaseous disruption of infalling objects and examine the extent of pre-processing in our simulations. In particular, we use the orbital histories of 132 427 (sub)haloes to identify where, when and how these objects lose their gas since crossing into the infall region, which we define as $1-4 R_{200}$. Our main conclusions are as follows:

\begin{itemize}
    \item On average, infalling objects lose nearly all of their gas around $1.7 R_{200}$ in the 3D perspective and $\sim R_{200}$ in the PROJ perspective. By constructing stacked fractional gas-loss trends as a function of cluster-centric distance, we show that in a 3D perspective objects lose $\sim 30$ per cent of their gas at $\sim 2R_{200}$ , where there is then a dramatic increase in gas-loss until $\sim 1.7R_{200}$. In the PROJ perspective, these trends are translated $\sim 0.7R_{200}$ closer to the cluster. Our results agree with and extend the work done in \citet{Arthur19}, by showing that, on average, objects actually lose their gas at a characteristic radius in the infall region. The increase of fractional gas-loss with decreasing cluster-centric distance is not due to contamination from outgoing objects, which agrees with \citet{Lotz19}. 

    \item By splitting the radial trends by subhalo status, we show that subhaloes lose their gas much further out than haloes. $\sim 60$ per cent of gas is depleted from subhaloes by $\sim 3R_{200}$ , whereas haloes only reach this level of disruption by $\sim 1.7R_{200}$. However, removing subhaloes residing in cluster-mass hosts in the infall region from the radial trends, we find that subhaloes in group-mass and low-mass hosts follow essentially the same radial gas-loss trends as haloes, suggesting that cluster-mass hosts environments are responsible for removing most of the gas observed for the whole subhalo sample. At present, the relative importance of gas disruption (either by to sub-dominant or cluster-mass hosts environments), secular processes or numerically enhanced stripping of subhaloes on these results is unclear.
    
    \item The phase-space analysis of the objects in the sample shows that the infalling branch is established within $\sim1-3$ Gyr since the objects entered the infall region. After this time, a characteristic gas-loss cluster-centric radius can be identified at $\sim 1.5R_{200}$. Around $4-5$ Gyr, the backsplash population branch becomes apparent. Independent of their time since they entered the infall region, any object within $\sim 1.5R_{200}$ has lost nearly all of its gas. On the other hand, for objects that remain outside this region, their gas-loss fraction is dependent on the time they spent in the infall region.
    
    \item Subhaloes lose their gas much quicker than haloes after entering the infall region of the main cluster halo. Within 1 Gyr, the subhalo population shows significant depletion of its gas. Generally haloes retain their gas for much longer, but there is a population of these objects that lose their gas within the first few Gyrs after infall as well. These are either backsplash haloes or haloes that have entered the characteristic radius at $\sim 1.5 R_{200}$, but are not yet classified as a subhalo of the cluster.
    
    \item Subhaloes lose their gas more quickly within higher mass host halo environments. In group-mass host haloes, the gas of subhaloes is typically depleted by $\sim 1.5$ Gyr since they became subhaloes, whilst in Milky Way like hosts, it is typically depleted by $\sim 4$ Gyr. Our results agree with those found in \citet{Bahe15,Roberts17}, and \citet{Cora19} who suggest that higher mass hosts quench satellites quicker than low mass hosts. However, when one only considers the first $\sim 20$ per cent of subhalo gas-loss, all host environments are nearly equally as efficient.
    
    \item Subhaloes in group and low-mass host haloes appear to gain stellar mass at high cluster-centric radii even after spending 4 Gyrs within host halo environments. On the other hand, on average, subhaloes of cluster-mass haloes do not form stars. In fact, we find that their stellar material slightly declines as a function of time spent within the host. We note that this decline might not be solely the result of a lack of stellar production: it is possible that these objects are indeed using their gas to form stars, but due to stripping or ejection processes, stellar material can be lost in subhaloes, outweighing star formation. This would result in a net stellar loss, despite the subhaloes actually forming stars. Disentangling the relative relevance of these processes would be a primary aim for future analysis. On the other hand, as the stellar component is significantly more bound than the gas in a halo, we find this scenario unlikely. Our results are in partial disagreement with \citet{Wetzel13}, who found that there is a delay in satellite quenching once they accrete onto their host, whereby all satellites are still allowed to form stars, and not only the ones in group and low-mass hosts. This mass-dependent stellar-loss trend, however, is compatible with the analysis presented in \citet{Cora19}, who pointed that secular processes play a major role on the decline of the star formation of high-mass galaxies, potentially allowing galaxies with lower mass to keep forming stars. 
    
    \item Lower mass subhaloes retain their gas for longer once they have accreted onto a host than higher mass subhaloes. This is easily explained by the fact that massive subhaloes are in more massive hosts, which as seen, deplete subhalo gas more efficiently, e.g. subhaloes in more massive hosts would accrete with larger velocities in the high density environment of their host, which increases the ram pressure they experience. These results suggest that gas depletion in our subhalo population is driven predominately by host mass rather than artificially enhanced numerical stripping found in recent studies \citep{vanDenBosch18a, vanDenBosch18b}.

\end{itemize}

In summary, we conclude that in the infall region, infalling objects suffer significant gaseous disruption that correlates with their time-since-infall and cluster-centric distance. Subhaloes suffer more disruption further out from the main cluster halo and quicker than the halo population, and we find that the gas-loss radial trend is dominated by subhaloes residing in cluster-mass hosts in the infall region. Considering only the subhaloes in group-mass and low-mass host halo environments, we find a similar radial trend to the halo population. We have shown that this is not solely driven by artificially enhanced stripping due to particle number resolution. However, as the spatial resolution of the simulation might induce an enhanced stripping of gas particles on our objects if their intrahalo medium is not fully resolved, we understand the characteristic radius found in our radial gas-loss trends as an upper bound of the accretion shock radius, rather than its actual value. 

Our results suggest that gaseous disruption in the infall region is a combination of pre-processing and cluster quenching (for subhaloes within cluster-mass host haloes), and object gas depletion at a characteristic radius that behaves as an accretion shock. However, some questions remain and could be the subject of a future study. For instance, a prime concern would be to investigate how environmental effects alone contribute to the gas-loss of infalling objects, by examining how (particle and mass) resolution and different subgrid physics affect our trends. Fortunately, \textsc{The Three Hundred} has been run with multiple codes, each containing different subgrid physics with different calibrations. Alongside this, there are $2-5$ clusters within the sample that have been re-run with eight times better mass resolution. By using this data, a follow-up study could relatively easily target the question of how dominant environment is. In addition to this, by disentangling the gaseous components in subhaloes into the cold inter stellar medium and the hot halo, a future study could pin down what kind of host environments are necessary in order to exclusively strip the hot component, essentially starving the satellite within the subhalo, and which environments are necessary in order to disturb the cold inter stellar medium.

\section*{Acknowledgements}

This work has been made possible by the `The Three Hundred' collaboration.\footnote{\url{https://www.the300-project.org}} The project has received financial support from the European Union's Horizon 2020 Research and Innovation programme under the Marie Sklodowskaw-Curie grant agreement number 734374, i.e. the LACEGAL project.

The authors would like to thank The Red Espa\~{n}ola de Supercomputaci\'{o}n for granting us computing time at the MareNostrum Supercomputer of the BSC-CNS where most of the cluster simulations have been performed. Part of the computations with \code{GADGET-X} have also been performed at the `Leibniz-Rechenzentrum' with CPU time assigned to the Project `pr83li'. We also thank the anonymous referee for their helpful comments, which improved the quality of the manuscript.

RM, AK and GY are supported by the MICIU/FEDER through grant number PGC2018-094975-C21. AK further acknowledges support from the Spanish Red Consolider MultiDark FPA2017-90566-REDC and thanks Metallica for ride the lightning. SAC acknowledges funding from {\it Consejo Nacional de Investigaciones Cient\'{\i}ficas y T\'ecnicas} (CONICET, PIP-0387), and {\it Universidad Nacional de La Plata} (G11-150), Argentina. KD acknowledges support by the Deutsche Forschungsgemeinschaft (DFG, German Research Foundation) under Germany's Excellence Strategy - EXC-2094 - 390783311.

The authors contributed to this paper in the following ways: RM, JA, FRP, MG and AK formed the core team. RM and JA analysed the data, produced the plots and wrote the paper along with FRP, MG, AK, WC, CW, and SAC.  GM, KD, and GY supplied the simulation data. All authors had the opportunity to provide comments on this work.

This work was created by making use of the following software: \code{Python}, \code{Matplotlib} \citep{Matplotlib-Hunter07}, \code{Numpy} \citep{Numpy-vanDerWalt11}, \code{scipy} \citep{Scipy-Virtanen19}, and \code{astropy} \citep{AstropyI,AstropyII}.

\section*{Data availability}

Data available on request to \textsc{The Three Hundred} collaboration, at \url{https://www.the300-project.org}.




\bibliographystyle{mnras}
\bibliography{archive}



\appendix


\bsp	
\label{lastpage}
\end{document}